\documentclass[10pt,aps,pra,twocolumn,showpacs]{revtex4}   
\usepackage{amsmath}    
\usepackage{amssymb}
\usepackage{graphicx}   

\usepackage{amscd}
\usepackage{subfigure}
\usepackage{epic}

\newcommand{\ri}{{ \rm i }}
\newcommand{\re}{{ \rm e }}
\newcommand{\rd}{{ \rm d }}

\newcommand{\ra}{{ \rm a }}
\newcommand{\rb}{{ \rm b }}
\newcommand{\DD}{\eta}

\newcommand{\be}{\begin{equation}}
\newcommand{\ee}{\end{equation}}

\usepackage{color}
\definecolor{blau}{rgb}{0,0,1}
\definecolor{gruen}{rgb}{0,1,0}
\definecolor{rot}{rgb}{1,0,0}
\definecolor{magenta}{rgb}{1,0,1}

\begin{document}

\title[Non-Hermitian quantum walks of ultracold Bosons]{Interaction-induced decoherence in non-Hermitian quantum walks of ultracold Bosons}
\author{K. Rapedius}
\author{H.~J. Korsch}
 \affiliation{FB Physik, TU Kaiserslautern, D-67653 Kaiserslautern, Germany}
 \email{rapedius@physik.uni-kl.de, korsch@physik.uni-kl.de}   

\date{\today}


\begin{abstract}
We study the decoherence caused by particle interaction for a conceptually simple model,
a quantum walk on a bipartite one-dimensional lattice with decay from every
second site. The corresponding non-interacting (linear) system 
has been shown to have a topological transition described by the average displacement
before decay. Here we use this topological quantity to distinguish
coherent quantum dynamics from incoherent classical dynamics caused by a breaking of the
translational symmetry. We furthermore analyze the behavior by means of
a rate equation providing a quantitative description of the 
incoherent nonlinear dynamics.
\end{abstract}

\maketitle

The quantum-to-classical transition often denoted as decoherence
is of increasing recent interest and considerable progress
has been achieved \cite{Joos03,Schl07}. 
Open many-particle systems, in particular
ultra-cold quantum gases and Bose-Einstein condensates, 
are well-suited for investigating decoherence processes in nonequilibrium quantum dynamics.
Recent examples of theoretical and experimental studies of such systems include quantum walks in optical lattices 
\cite{Kars09,Witt10,Chan11} and the decay dynamics in open, non-hermitian systems 
\cite{Schl06a,Schl06b,08nlLorentz,09ddshell,10nlret,Gebh12,Trim11,Witt08}, some of which explicitly address decoherence \cite{Gebh12,Trim11,Witt08}.  

\begin{figure}[htb]
\begin{center}

\includegraphics[width=0.43\textwidth]{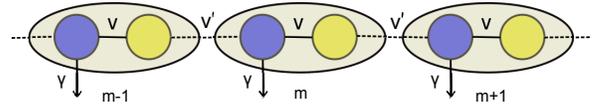}

\caption{\label{fig:bild}(Color online) Bipartite 1D lattice with couplings $v$ and $v'$
and decay from every second site with decay rate $\gamma$.}
\end{center}
\end{figure}

In this brief report, we analyze the mechanism of decoherence in the nonequilibrium dynamics of an interacting Bose gas by means of a
topological quantity introduced in \cite{Rudn09}. Topological properties have also been of interest in other recent investigations 
with cold atoms concerning, e.g., quantum Hall effects \cite{Hafe07,Gold11}, Berry phases \cite{Arik10,10expo}
or dissipative quantum wires \cite{Dieh11}.
We will demonstrate an interaction induced decoherence, i.e.~a quantum-to-classical transition for a conceptually simple model
system.
In particular, we study the quantum walk of ultracold bosons in a deep bipartite 1D 
optical lattice with two different tunneling rates and decay from every second 
site described in a non-hermitian tight-binding model
considered by Rudner and Levitov \cite{Rudn09} in their analysis of a topological
transition. Here we generalize and study    
the non-hermitian discrete Gross-Pitaevskii (or nonlinear Schr\"odinger) equations 
\begin{equation}\begin{split}\label{psiA}
\ri \hbar \dot a_m &= \left(\epsilon_\ra+ g |a_m^2|\right) a_m- \frac{v}{2} b_m-\frac{v'}{2}b_{m+1} \\ 
 \ri \hbar \dot b_m &= \left(\epsilon_\rb- \ri \frac{\gamma}{2} + g |b_m|^2\right) b_m- \frac{v}{2}a_m-\frac{v'}{2}a_{m-1}
\end{split}\end{equation}
where the nonlinear terms $g |a_m|^2$, $g |b_m|^2$ model the interaction between the particles.
(Note that such an equation introduced heuristically in studies of ultra-cold bosonic gases or
optical devices \cite{Such11}
can be derived rigorously as a mean-field approximation from a multi-particle
Bose-Hubbard model with dissipation in the limit of high particle numbers in the system \cite{Trim11}. 
In a different approach, however, a mean-field limit with a
renormalized interaction term appears \cite{08nhbh_s,10nhbh}.)
The system (\ref{psiA}) is illustrated graphically in Fig.~\ref{fig:bild}.
Initially all particles occupy a single non-decaying site which corresponds to the initial conditions $a_m(0)=\delta_{m0}$, $b_m(0)=0$ at time $t=0$.

In the noninteracting (linear) case, (\ref{psiA}) with $g=0$, Rudner and Levitov 
\cite{Rudn09} analyzed the topological quantity
\be
   \Delta m =\sum_m m P_m,  \quad P_m=\int_0^\infty \gamma |b_m(t)|^2 dt
\label{Delta_m}
\ee
describing the average displacement of the particles before their decay. By means of a Fourier
transformation to momentum space
they showed that $\Delta m$ is equal to the winding number of the relative phase between components of the Bloch wave function 
which leads to a quantization of $\Delta m$ as a function of the ratio $v/(v+v')$ of the tunneling coefficients: 
$\Delta m=1$ for $v/(v+v')<1/2$ whereas  $\Delta m=0$ for $v/(v+v')>1/2$. 
In \cite{Rudn09} the quantization of $\Delta m$ was shown 
to be robust against intersublattice dephasing and classical noise on the energy levels $\epsilon_a$, $\epsilon_b$ by means of 
numerical simulations. This quantized behavior was compared to the continuous dependence of $\Delta m$ on $v$ and $v'$ in the 
case of incoherent tunneling.

\begin{figure}[htb]
\begin{center}
\includegraphics[width=0.45\textwidth, height=0.3\textwidth] {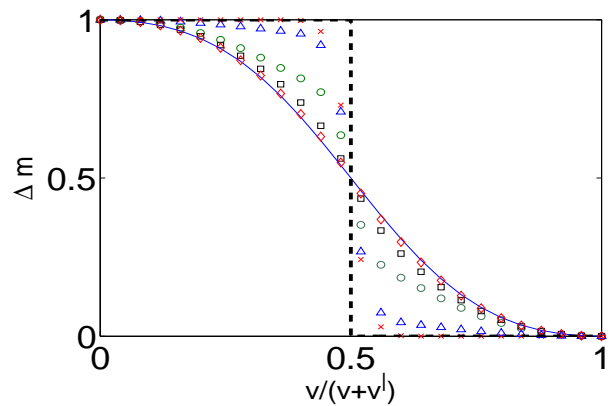}
\caption{\label{fig:Dm} (Color online) Displacement $\Delta m$ as a function of the ratio $v/(v+v')$ of the tunneling coefficients. '- -': 
linear ($g=0$) infinitely extended system, '{---}' incoherent hopping according to (\ref{Delta m}). The symbols represent numerical 
results for a finite system with $46$ lattice sites and periodic boundary conditions: '{$X$}': $g=0$, '{$\Delta$}': $g=0.2$,
'{$\circ$}': $g=0.5$, '$\square$': $g=1$ ,'{$\Diamond$}': $g=4$. The parameters in the numerical simulation 
are $\gamma=2$, $v'=0.5$ with $0 \le v <0.5 $ for $v/(v+v')<0.5$ and $v=0.5$ with $0 \le v' <0.5 $ for $v/(v+v')>0.5$, $\epsilon_a=0=\epsilon_b$.
}
\end{center}
\end{figure}

The latter can be obtained by assuming a randomly fluctuating phase between the sites so that
the rates for incoherent hopping are given by perturbation theory to the lowest order as
$\Gamma \propto v^2$, $\Gamma' \propto v'^2$ \cite{Van55}. Initially only the site $a_0$ is occupied.
After the very first hopping process to one of the decaying neighbour sites $b_0$ or $b_1$ corresponding to $m=0$ and $m=1$ respectively, 
the incoherent dynamics between the decaying sites $b_m$ is symmetric with regard to transport to the left and right so that the 
contributions of all subsequent hoppings to the displacement (\ref{Delta_m}) cancel out. Since the $m=0$ term does not contribute, 
the displacement $\Delta m$ is given by the probability of an initial hopping to site $b_1$ with $m=1$ which is equal to the 
corresponding relative hopping coefficient, i.e.
\be 
   \Delta m_{\rm incoherent} = \Gamma'/(\Gamma+\Gamma')=v'^2/(v^2+v'^2) \,.
   \label{Delta m}
\ee
Figure \ref{fig:Dm} compares the the quantized behavior of $\Delta m$ for coherent tunneling (dashed line) with 
the continuous curve (\ref{Delta m}) (solid line) for incoherent tunneling.  

In order to investigate the influence of a finite mean-field interaction $g \ne 0$ we integrate the system (\ref{psiA})
numerically on a finite lattice with periodic boundaries and parameters $\epsilon_a=\epsilon_b=0$. 
In all numerical calculations scaled units with $\hbar=1$ are used throughout this paper.
The results are also displayed in Fig.~\ref{fig:Dm}. 
Even in the linear case $g=0$ the numerical results for $\Delta m$ deviate from the perfectly quantized 
behavior of the idealized infinitely extended system around $v/(v+v')\approx 0.5$ due to effects of finite size and finite integration 
time. These effects were discussed in detail in \cite{Rudn09} and are not significantly altered by the mean-field interaction. 
For increasing values of the interaction parameter $g$, the values of 
$\Delta m$ deviate more and more from the quantized behavior of the linear system, approaching the incoherent tunneling curve 
for higher values of $g$. Thus the mean-field interaction, unlike other kinds of disturbances mentioned above, induces 
decoherence in the system.

\begin{figure}[htb]
\begin{center}
\includegraphics[width=0.45\textwidth, height=0.3\textwidth] {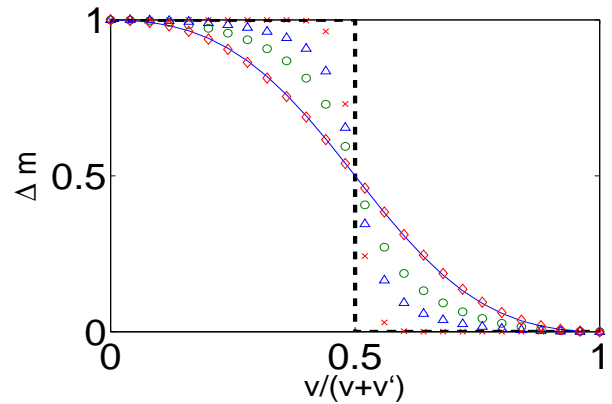}
\caption{\label{fig:Dm2}(Color online) As Fig.~\ref{fig:Dm} but for the linear 
system ($g=0$) with an energy offset $\DD$ of the initial site. '{$X$}': $\DD=0$, '{$\Delta$}': $\DD=0.05$,
'$\circ$': $\DD=0.1$, '{$\Diamond$}': $\DD=0.6$.
}
\end{center}
\end{figure}

The reason for this decoherence lies in the fact that the nonlinear interaction breaks the translational symmetry within the
non-decaying sublattice as will become clear in the following. To support our argument, we will demonstrate that, instead of 
considering interactions, the coherence observed in the linear system can also be disturbed by means of a much simpler symmetry 
breaking mechanism, namely by adding a constant energy shift to the initial site $a_0$. Figure \ref{fig:Dm2} shows the numerically calculated 
displacement $\Delta m$ as a function of $v/(v+v')$ for the system (\ref{psiA}) with $g=0$ and an additional energy 
shift $\DD$ of the initial site energy, i.e.~we make the replacement $\epsilon_a \rightarrow \epsilon_a +\DD \delta_{m0}$ in 
(\ref{psiA}). We observe that the values of $\Delta m$ more and more approach the incoherent hopping curve
for increasing values of the energy shift $\DD$. By means of a constant energy shift of a single site we are thus able to 
mimic the decoherence inducing effect of the mean-field interaction, thereby identifying the breaking of the translational symmetry
within the non-decaying sublattice as the main cause of decoherence. 

To gain further insight we analyze the time-dependence of the correlations $a_n^*(t)a_m(t)$, $b_n^*(t)b_m(t)$ and $b_n^*(t)a_m(t)$   
between the sites within and between the two sublattices which is most conveniently done by reexpressing our system by means of
a density matrix. 
At first we concentrate on the simpler case of a non-interacting system with an additional shift of the initial site. For convenience, we first rewrite the system (\ref{psiA}) with $g=0$ and an additional energy shift $\DD$ of the site 
$a_0$ in the compact form
\be
   \ri \hbar \dot c_\alpha = -J'_\alpha c_{\alpha+1}-J_\alpha c_{\alpha-1}+ E_\alpha c_\alpha
\label{dot c}
\ee
where
\be
   c_\alpha=\left\{ \begin{array}{cl}
		a_{m=\alpha/2}, & \alpha\ \text{even}\\
	        b_{m=(\alpha+1)/2}, & \alpha \ \text{odd}
		\end{array}
        \right. , 
 E_\alpha=\left\{ \begin{array}{cl}
		\epsilon_\ra + \DD \delta_{\alpha0}, & \alpha \ \text{even}\\
		\epsilon_\rb- \ri \gamma/2, & \alpha \ \text{odd}
		\end{array}
        \right.
\ee
and
\be
   J_\alpha=\left\{ \begin{array}{cl}
		v/2, & \alpha \ \text{even}\\
		 v'/2, & \alpha \, \text{odd}
		\end{array}
        \right. \, ,\quad  
   J'_\alpha=\left\{ \begin{array}{cl}
		v'/2, & \alpha \ \text{even}\\
		 v/2, & \alpha \, \text{odd}
		\end{array}
        \right. \,.
\ee
For the dynamics of the density matrix elements $\rho_{\alpha\beta}=c_\alpha^*c_\beta$ 
Eq.~(\ref{dot c}) then yields
\begin{eqnarray}
  \ri \hbar \dot \rho_{\alpha\beta} &=& (E_\beta-E_\alpha^*) \rho_{\alpha\beta} + J'_\alpha \rho_{\alpha+1,\beta}+ J_\alpha \rho_{\alpha-1,\beta} \nonumber \\
                                      &&- J'_\beta \rho_{\alpha,\beta+1}- J_\beta \rho_{\alpha,\beta-1} \,.
\label{dot rho}
\end{eqnarray}
The initial conditions at $t=0$ read
$\rho_{\alpha \beta}(0)=\delta_{\alpha \beta} \delta_{\alpha 0}$. As above we assume $\epsilon_a=0=\epsilon_b$ in the following.
We consider the dynamics of the central diagonal element of the density matrix
\be
   \ri \hbar \dot \rho_{00}=J'_0(\rho_{10}-\rho_{01})+J_0(\rho_{-1,0}-\rho_{0,-1})
   \label{dot rho00}
\ee
which depends on the nearest off-diagonal elements. As an example we have a closer look at the element
\be
   \ri \hbar \dot \rho_{01}=(E_1-E_0^*)\rho_{01}+J'_0 \rho_{11}+J_0\rho_{-1,1}-J'_1\rho_{02}-J_1\rho_{00} \,.
\ee
If we neglect the off-diagonal elements between two odd (i.e.~decaying) sites we arrive at
\be
   \ri \hbar \dot \rho_{01}\approx (E_1-E_0^*)\rho_{01}+J'_0 (\rho_{11}-\rho_{00})-J_0\rho_{02} 
 \label{dot rho02_app}
\ee
where we have used $J_1=J'_0$.
The dynamics of the off-diagonal element $\rho_{02}$ must be examined more closely. Since our initial state is localized at site $0$ we can neglect
decaying off-diagonal elements not involving site $0$. If we additionally neglect correlations over a distance of three or more sites 
Eq.~(\ref{dot rho}) yields $\ri \hbar \dot \rho_{02}\approx (E_2-E_0^*)\rho_{02}-J_0\rho_{01}$. Assuming a slow time-dependence of $\rho_{01}$ compared to 
$\hbar/(E_2-E_0^*)$, we obtain $\rho_{02} \approx \rho_{01}(1-\exp(-i (E_2-E_0^*)t/\hbar))J_0/(E_2-E_0^*)$. For $|E_2-E_0^*|=|\DD| \gg J_0$ the 
term $J_0\rho_{02}$ in (\ref{dot rho02_app}) can thus be neglected. Physically this corresponds to a suppression of resonant tunneling 
between the sites $0$ and $1$ due to the large difference $|E_2-E_0^*|=|\DD|$ of the on-site energies.
Following the reasoning in \cite{02emission} we assume a slow time-dependence of $\rho_{00}$ and $\rho_{11}$ compared to the real part of $\hbar/(E_1-E_0^*)$, 
which is justified for $ |{\rm Re}(E_1-E_0^*)|=|\DD| \gg J_0$, to obtain 
\be
   \rho_{01}=\frac{J'_0}{\hbar} \frac{\rho_{11}-\rho_{00}}{\DD+\ri \gamma/2}=\rho_{10}^*
\ee 
with $E_1-E_0^*=-\DD-\ri \gamma/2$ and, analogously $\rho_{0,-1}=(J_0/\hbar)(\rho_{-1,-1}-\rho_{00})/(\DD+\ri \gamma/2)=\rho_{-1,0}^*$.
Thus the dynamics of the central diagonal matrix element (\ref{dot rho00}) becomes 
\be
   \dot \rho_{00} \approx -(\Gamma+\Gamma') \rho_{00}+\Gamma \rho_{-1,-1}+\Gamma' \rho_{11}
   \label{rate00}
\ee
where 
\be
   \Gamma=\frac{(J_0/\hbar)^2\gamma}{\DD^2+\gamma^2/4}=\frac{(v/\hbar)^2\gamma}{4\DD^2+\gamma^2},\,
   \Gamma'=\frac{(J'_0/\hbar)^2\gamma}{\DD^2+\gamma^2/4}=\frac{(v'/\hbar)^2\gamma}{4\DD^2+\gamma^2}.
\ee
Note the quadratic dependencies $\Gamma \propto v^2$, $\Gamma' \propto v'^2$ already stated before Eq.~(\ref{Delta m}).
For a sufficiently strong decay coefficient $\gamma$ (compared to the tunneling coefficients $v$ and $v'$) one can assume that the correlations between sites further away 
from site $0$, which vanish at $t=0$, do not build up significantly in the course of the decay process. As an approximation we 
may thus assume an incoherent dynamics as described by (\ref{rate00}) for all sites with some local site-dependent transition 
rates $\Gamma_\alpha$ and $\Gamma_\alpha'$ yielding a rate equation
\begin{eqnarray}
 \dot \rho_{\alpha \alpha} \approx -(\Gamma+\Gamma') \rho_{\alpha \alpha} \qquad \qquad\qquad \qquad \qquad \qquad \qquad \label{rate2} \\
\quad +\left\{ \begin{array}{cl}
		\Gamma_\alpha \rho_{\alpha -1, \alpha -1}+\Gamma_\alpha' \rho_{\alpha+1,\alpha+1} , & \alpha \ \text{even}\\[2mm]
		-\gamma \rho_{\alpha\alpha}+\Gamma_\alpha' \rho_{\alpha -1, \alpha -1}+\Gamma_\alpha \rho_{\alpha+1,\alpha+1}, & \alpha \ \text{odd}.
		\end{array} 
        \right.  \nonumber
\end{eqnarray}
However, due to the decay in the system the quantities that we are interested in, namely $\Delta m$ and the 
occupation of the central site $\rho_{00}$ (cf.~below) are not sensitive to the exact dynamics in the outer sites whose main effect 
in this context is the suppression of boundary effects. Thus for simplicity we make the approximation of constant transition rates 
$\Gamma_\alpha=\Gamma$ and $\Gamma_\alpha'=\Gamma'$ in the following.
Within this approximation, equation (\ref{rate2}) can be solved in closed form in Fourier space with the result (\ref{Delta_m})
for $\Delta m$ (see Appendix).

\begin{figure}[htb]
\begin{center}
\includegraphics[width=0.47\textwidth, height=0.2\textwidth] {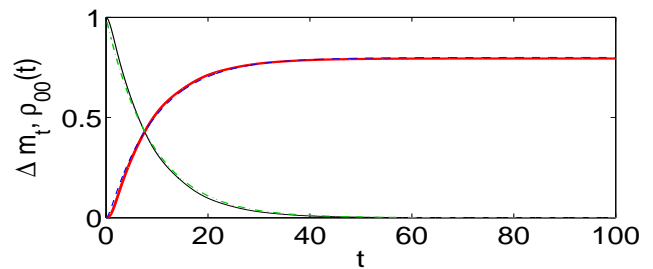}
\caption{\label{fig:Dyn} (Color online) (Color online)  Dynamics of the linear system ($g=0$) with an energy offset $\DD=0.6$ of the initial site $\alpha=0$ for the parameters $v=0.25$, $v'=0.5$, $\gamma=2$ and $\epsilon_a=0=\epsilon_b$.
The numerical integration of the full system (\ref{dot c}) ($\rho_{00}$:solid line, $\Delta m_t$: bold solid line) is compared with the dynamics according to the rate 
equations (\ref{rate2}) ($\rho_{00}$:dashed dotted line, $\Delta m_t$: dashed line) . 
}
\end{center}
\end{figure}

\begin{figure}[htb]
\begin{center}
\includegraphics[width=0.47\textwidth, height=0.2\textwidth] {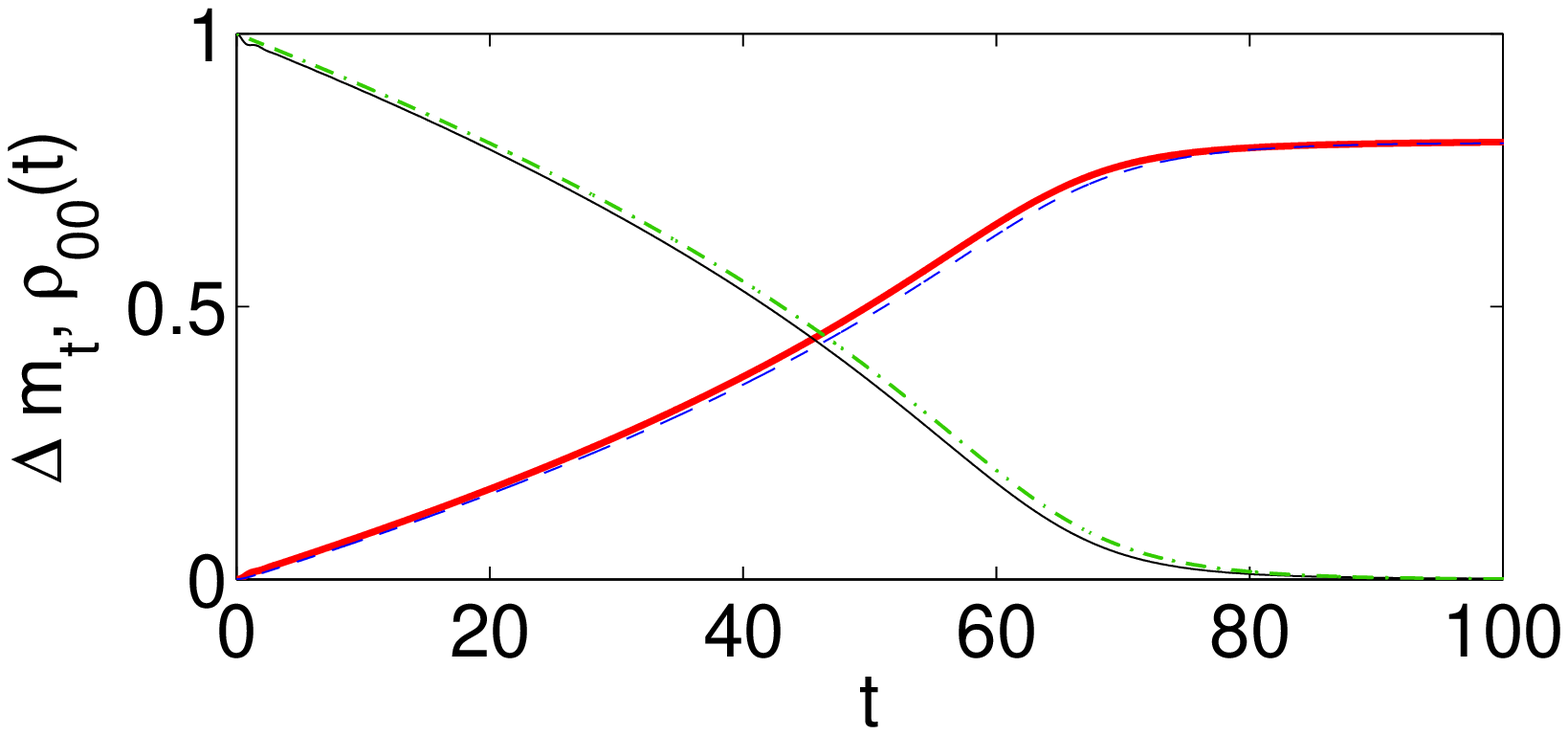}
\includegraphics[width=0.47\textwidth, height=0.2\textwidth] {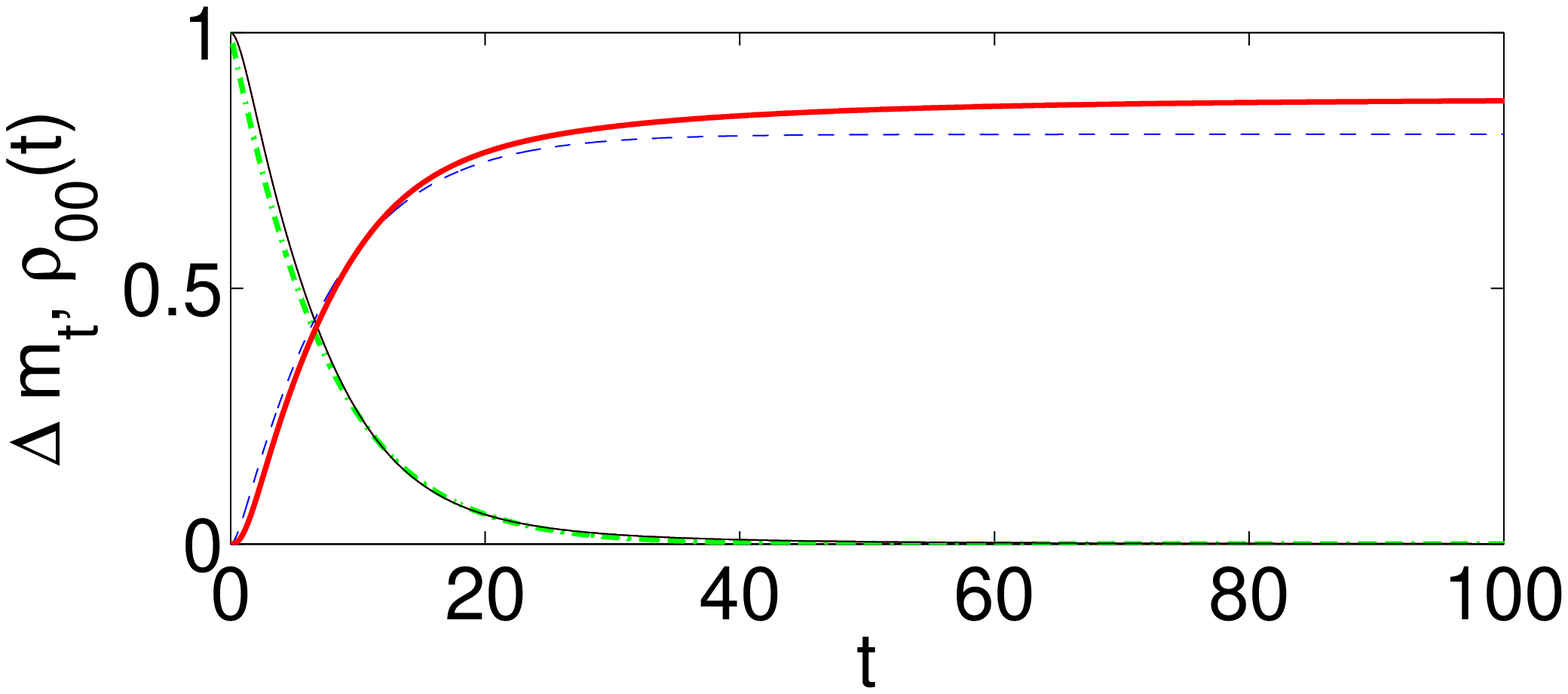}
\caption{\label{fig:Dyn2} (Color online) As Fig.~\ref{fig:Dyn} but for the nonlinear system with $g=4$ (upper panel) and $g=0.5$ (lower panel).
}
\end{center}
\end{figure}


In order to compare the dynamics obtained from the rate equation (\ref{rate2}) with the dynamics of the full system 
we define the time-dependent displacement $\Delta m_t$, which is obtained if the time integration in (\ref{Delta m}) is only
performed up to a finite time $t$ such that $\Delta m ={\rm lim}_{t \rightarrow \infty} \Delta m_t$. 
An example of the dynamics of $\Delta m_t$ for $\DD=0.6$ is shown in Fig.~\ref{fig:Dyn} together
with the corresponding decay of the central site occupation $\rho_{00}=|a_0|^2$. While $\rho_{00}$ decays 
exponentially, the displacement $\Delta m_t$ increases until it reaches its final value, given approximately by
(\ref{Delta m}), for long times. For both quantities the numerically exact calculation is reasonably well approximated by the rate
equation result.

Let us now return to the original, nonlinear problem with a finite interaction $g \ne 0$. The role of the energy offset $\DD$ 
is now played by the mean-field interaction. For a sufficiently strong decay coefficient $\gamma$, the instantaneous energy offset
between the central site and its neighbors is approximately given by the local mean-field interaction term at site $0$. To obtain 
an approximate description of the nonlinear system dynamics we thus use the rate equation (\ref{rate2}) with the replacement 
$\DD=g \rho_{00}(t)$. Even though the hopping rates $\Gamma$ and $\Gamma'$ are now time-dependent, their ratio remains constant
on this level of approximation such that the final displacement $\Delta m$ is still well approximated by (\ref{Delta m}).
The upper panel of Fig.~\ref{fig:Dyn2} demonstrates that for $g=4$, corresponding to the incoherent regime (cf.~Fig.~\ref{fig:Dm}), the modifications of the system dynamics due to the nonlinearity are well described by
the effective rate equation, both for the time-dependent displacement $\Delta m_t$ and the central site occupation 
$\rho_{00}$. The latter now shows a nonexponential decay behavior typical of open nonlinear 
systems \cite{Schl06a,Schl06b,08nlLorentz, 09ddshell}. In the long time limit, the exponential decay of the linear system is 
recovered as the influence of the nonlinear interaction term $g \rho_{00}(t)$ becomes negligible. 
For comparison the lower panel of Fig.~\ref{fig:Dyn2} shows the dynamics for a smaller interaction $g=0.5$ corresponding to an intermediate, still 
partially coherent regime (cf.~Fig.~\ref{fig:Dm}). While the decay of the central site occupation is still described 
reasonably well by the rate equation approach we find clear deviations for the time-dependent displacement $\Delta m_t$ 
as expected.

In summary, the interaction-induced decoherence of cold bosons spreading in a bipartite optical lattice with decay 
was analyzed by means of a topological quantity. Using numerical and analytical methods, the breaking of translational symmetry within the non-decaying sublattice
and the resulting suppression of resonant tunneling was identified as the cause of decoherence. For the regime of strong decay and strong interaction, a rate equation providing a 
quantitative description of the system's incoherent nonlinear dynamics was derived. The authors hope that the present model 
study can contribute to a better understanding of the transition from coherent quantum dynamics to incoherent classical dynamics in mesoscopic systems.

We thank Eva-Maria Graefe and Dirk Witthaut for useful comments and suggestions.

\section*{Appendix}
\label{app}
Changing the notation to
$p_m^{(\pm)}=\rho_{\alpha \alpha}$ for $\alpha$ even (+) or odd (-) 
 Eq.~(\ref{rate2}) with constant decay rates $\Gamma_\alpha=\Gamma$ and $\Gamma_\alpha'=\Gamma'$ reads
\begin{eqnarray}
\dot p_m^{(+)}&=&-\Gamma_0\,p_m^{(+)}+\Gamma p_m^{(-)}+\Gamma' p_{m+1}^{(-)}
\nonumber\\
\dot p_m^{(-)}&=&-\Gamma_0'\,p_m^{(-)}+\Gamma' p_{m-1}^{(+)}+\Gamma p_{m}^{(+)}
\label{rateA1}
\end{eqnarray}
with $\Gamma_0=\Gamma+\Gamma'$ and  $\Gamma_0'=\Gamma_0+\gamma$.
As in \cite{Rudn09} we switch to the momentum representation
\be
p_m^{(\pm)}=\tfrac{1}{2\pi}\oint \rd k\,\re^{\ri km}q_k^{(\pm)}
\ , \quad 
q_k^{(\pm)}=\sum_m\re^{-\ri km}p_m^{(\pm)}\,,
\ee
where the integration extends over the Brillouin zone $-\pi \le k<\pi$.
The resulting equations for the momentum distributions
\begin{eqnarray}
\dot q_k^{(+)}&=&-\Gamma_0q_k^{(+)}+\Gamma_k q_k^{(-)}\nonumber\\
\dot q_m^{(-)}&=&-\Gamma_0'q_k^{(-)}+\Gamma_k^* q_k^{(+)}
\label{rateA2}
\end{eqnarray}
with $\Gamma_k=\Gamma+\Gamma'\,\re^{\ri k}$ can be solved immediately
and a solution for the 
initial conditions  $q_k^{(+)}(0)=1$\,, \ $q_k^{(-)}(0)=0$
(the translation of $p_m^{(+)}(0)=\delta_{m0}$\,,\ $p_m^{(-)}(0)=0$)
is
\begin{eqnarray}
q_k^{(+)}&=&\frac{1}{\lambda_+-\lambda_-}\,\Big(
(\lambda_++\Gamma_0')\,\re^{\lambda_+t}-(\lambda_-+\Gamma_0')\,\re^{\lambda_-t}
\Big)\nonumber\\
q_k^{(-)}&=&\frac{\Gamma_k^*}{\lambda_+-\lambda_-}\,\Big(
\re^{\lambda_+t}-\re^{\lambda_-t}\Big)
\label{ratesol}
\end{eqnarray}
with
\be
\lambda_\pm=-(\Gamma_0+\Gamma_0')/2 \pm\sqrt{\gamma^2/4+|\Gamma_k|^2}\,.
\ee
The momentum representation of (\ref{Delta_m}) is given by
\begin{eqnarray}
   \Delta m &=&\gamma \int_0^\infty\!d t\,\sum_m mp_m^{(-)}(t)
= \gamma \int_0^\infty  dt\, \ri \partial_kq_k^{(-)}(t)\big|_{k=0}\nonumber\\
 &=&\ri \gamma\,\partial_k Q_k\big|_{k=0} \quad \textrm{with}\quad \
 Q_k=\int_0^\infty\!d t\,q_k^{(-)}(t)\,.
\label{Delta_m-A}
\end{eqnarray}
Integration of the solution $q_k^{(-)}(t)$ in (\ref{ratesol}) yields
\be
Q_k=\frac{\Gamma_k^*}{\Gamma_0\Gamma_0'-|\Gamma_k|^2}
\quad \textrm{and}\quad \
\Delta m=\frac{\Gamma'}{\Gamma+\Gamma'}\,.
\ee



\end{document}